# Predicting Oxide Glass Properties with Low Complexity Neural Network and Physical and Chemical Descriptors

Suresh Bishnoi[1], Skyler Badge[2], Jayadeva[2,*], N. M. Anoop Krishnan[3,4,*]

[1]School of Interdisciplinary Research, Indian Institute of Technology Delhi, Hauz Khas, New Delhi 110016, India
[2]Department of Electrical Engineering, Indian Institute of Technology Delhi, Hauz Khas, New Delhi 110016, India
[3]Department of Civil Engineering, Indian Institute of Technology Delhi, Hauz Khas, New Delhi 110016, India
[4]School of Artificial Intelligence, Indian Institute of Technology Delhi, Hauz Khas, New Delhi 110016, India

*Corresponding authors: jayadeva@iitd.ac.in, krishnan@iitd.ac.in

**Abstract**
Due to their disordered structure, glasses present a unique challenge in predicting the composition–property relationships. Recently, several attempts have been made to predict the glass properties using machine learning techniques. However, these techniques have the limitations, namely, (i) predictions are limited to the components that are present in the original dataset, and (ii) predictions towards the extreme values of the properties, important regions for new materials discovery, are not very reliable due to the sparse datapoints in this region. To address these challenges, here we present a low complexity neural network (LCNN) that provides improved performance in predicting the properties of oxide glasses. In addition, we combine the LCNN with physical and chemical descriptors that allow the development of universal models that can provide predictions for components beyond the training set. By training on a large dataset (~50000) of glass components, we show the LCNN outperforms state-of-the-art algorithms such as XGBoost. In addition, we interpret the LCNN models using Shapely additive explanations to gain insights into the role played by the descriptors in governing the property. Finally, we demonstrate the universality of the LCNN models by predicting the properties for glasses with new components that were not present in the original training set. Altogether, the present approach provides a promising direction towards accelerated discovery of novel glass compositions.

**Introduction**
Glasses are archetypical disordered materials that are formed by the fast quenching of liquids (Varshneya and Mauro, 2019). The disordered structure of glasses makes it extremely challenging to understand composition–structure–property relationships (Varshneya and Mauro, 2019). Despite the ubiquitous use of glasses in applications ranging from windshields, kitchen-wares, optical fibres, and display screens, to more specialized ones such as nuclear waste immobilization (Delaye et al., 2011) and biomedical implants (Bhaskar et al., 2020; Wilson et al., 1981), the glass design still relies on a trial-and-error approach, where intuition and domain expertise play key roles. Developing a comprehensive understanding of glass properties in terms of more fundamental features of their ingredients such as atoms and compounds, that is, *decoding the glass genome*, has been proposed a key problem for accelerating glass discovery (Mauro, 2018; Mauro et al., 2016).

Recently, machine learning (ML) approaches have been widely used for accelerating materials discovery (Meredig et al., 2014; Ramprasad et al., 2017; Schmidt et al., 2019; Ward et al., 2016). Specifically, in glasses, ML has been used for predicting several properties of glasses such as optical, physical, mechanical, and electrical properties (Bishnoi et al., 2019; Cassar et al., 2018, 2021b; Deng, 2020; Ravinder et al., 2020, 2021; Yang et al., 2019). While most of these approaches use composition as an input, some consider additional parameters such as testing conditions or synthesis conditions also to obtain improved predictions of glass properties (Cassar, 2021; Han et al., 2020; Krishnan et al., 2018; Liu et al., 2019; Zaki et al., 2021; Zhang et al., 2020). In addition, to gain insights into the composition–property relationships, interpretable algorithms such as Shapely additively explanations have been proven to be useful (Alcobaça et al., 2020; Cassar et al., 2021a; Ravinder et al., 2021; Zaki et al., 2022). Specifically, they allow domain experts to decode the role of the network formers, modifiers and intermediates present in a glass compositions in controlling a given property. In addition, they also provide insights into the coupled effects, that is the effect of one component on other components present in a glass, through SHAP interaction values (Ravinder et al., 2021). Some examples of such coupled effects in glasses include the boron anomaly, mixed modifier effects and the Loewenstein rule – or its violation.

While such models are useful for understanding the role of individual components, they suffer from several limitations, including : (i) Models are limited to components on which they have been trained. They cannot be used on a glasses with newer components, e.g. a model trained on sodium silicate glasses cannot be used to predict the properties of sodalimesilicate glasses. (ii) Glass properties are governed by those of their components, which are, in turn, governed by their atomic and electronic structures. Composition based models do not account for these features, and hence give no insights on their roles governing a property. (iii) Since these models are data-driven in nature, they train well in regions with abundant data. In regions with sparse or extreme values of a property, models tend to exhibit high variance.

To address these challenges, several physics and chemistry based descriptors have recently been proposed, along with the use of machine learning, to predict glass properties such as viscosity, density and elastic modulus (Cassar, 2021; Hwang et al., 2020; Liu et al., 2019; Shi et al., 2020). These models provide a promising route toward understanding the *glass genome*. However, all these machine learning (ML) models still exhibit the major limitation mentioned earlier, when used for prediction on new compositions. The quality of the ML model in a domain significantly depends on the availability of high quality data in that region. ML models are notorious for performing poorly in sparse data regions. Typically, in materials we are interested in exploring new materials in a region where there is sparce data. For example, glass compositions with maximum hardness, minimum thermal expansion, minimum loss, will all be located in regions where data is sparse. As such, while the overall ML model may be exhibiting good performance, the performance of the model in the specific extreme regions, where the data is sparse may be poor. Reliable use of ML models for accelerated materials discovery requires algorithms that give high accuracy for all regions of training domain irrespective of whether the data in that region is sparse or not.

Here, we propose a low-complexity neural network (LCNN) algorithm, which is used to predict nine properties of oxide glasses in terms of their physical and chemical descriptors. We show that the LCNN with a weighted loss function can provide high accuracy predictions in all the regions of the training data, including the ones where the data is sparse. Further, we show that the LCNN exhibits excellent property prediction, giving better performance than SOTA tree-based algorithms such as extreme gradient boosting (XGBoost). Finally, we show that the model developed based on the physical and chemical descriptors can be used to reliably predict the properties of glass compositions with new components that are unseen by the model.

**RESULTS**
**Dataset and feature engineering**
The dataset used in this work consists of ~50,000 glass compositions with 34 oxide components and nine properties. This dataset is from previous works Refs. (Bishnoi et al., 2021; Ravinder et al., 2020). Details of dataset preparation and cleaning can be found therein (see Methodology). The 34 input oxide components were transformed into handcrafted twelve descriptors (D1, D2, …, D12) as shown in Table 1. These twelve descriptors were used as the input for predicting the nine properties of glasses. Note that the descriptors developed were based on domain knowledge considering the following key points. (i) Properties of oxide glasses are highly dependent on the percentage of oxygen, network formers and modifiers. (ii) Valence of the former and modifier play crucial roles in determining the cationic strength, and hence the glass property. (iii) Atomic properties such as mass, number, radii govern the packing of the glass structure, which affects other properties. (iv) Finally, the polarizability of the glass govern the local charge distribution, which also is known to affect material properties. It is worth noting that in comparison to previous brute force ML approaches, the number of descriptors used in the present work is significantly lower, making the model parsimonious and potentially improving the interpretability.

**Table 1.** Descriptors generated from the mole percentage composition of oxides. An example case of $(CaO)_x(Al_2O_3)_y(SiO_2)_z$ with x, y, z as mol % (that is, x + y + z = 100) is shown.

| Code | Descriptors | Description |
|---|---|---|
| D1 | Oxygen percentage (O %) | $O\% = \dfrac{x + 3y + 2z}{2x + 5y + 3z}$ |
| D2 | Network former percentage (NF %) | $NF\% = \dfrac{2y + z}{2x + 5y + 3z}$ |
| D3 | Network modifier percentage (NM %) | $NM\% = \dfrac{x}{2x + 5y + 3z}$ |
| D4 | Valency of network former (VNF) | $VNF = y(3*2) + z(4)$ |
| D5 | Valency of network modifier (VNM) | $VNM = x(2)$ |
| D6 | Atomic mass(u) | $x(M_{Ca} + M_O) + y(2*M_{Al} + 3*M_O) + z(M_{Si} + 2*M_O)$ |
| D7 | Atomic volume (Å³) | $x(R_{Ca}^3 + R_O^3) + y(2*R_{Al}^3 + 3*R_O^3) + z(R_{Si}^3 + 2*R_O^3)$ |
| D8 | Atomic Number | $x(Z_{Ca}) + y(2*Z_{Al}) + z(Z_{Si})$ |
| D9 | Atomic radius (Å) | $x(R_{Ca} + R_O) + y(2*R_{Al} + 3*R_O) + z(R_{Si} + 2*R_O)$ |

| D10 | Van der waal radius (Å) | $x(R_{Ca} + R_O) + y(2 * R_{Al} + 3 * R_O) + z(R_{Si} + 2 * R_O)$ |
| D11 | Covalent radius (Å) | $x(R_{Ca} + R_O) + y(2 * R_{Al} + 3 * R_O) + z(R_{Si} + 2 * R_O)$ |
| D12 | Dipole polarizability (a.u.) | $x(\alpha_{Ca} + \alpha_O) + y(2 * \alpha_{Al} + 3 * \alpha_O) + z(\alpha_{Si} + 2 * \alpha_O)$ |

**LCNN based property prediction**

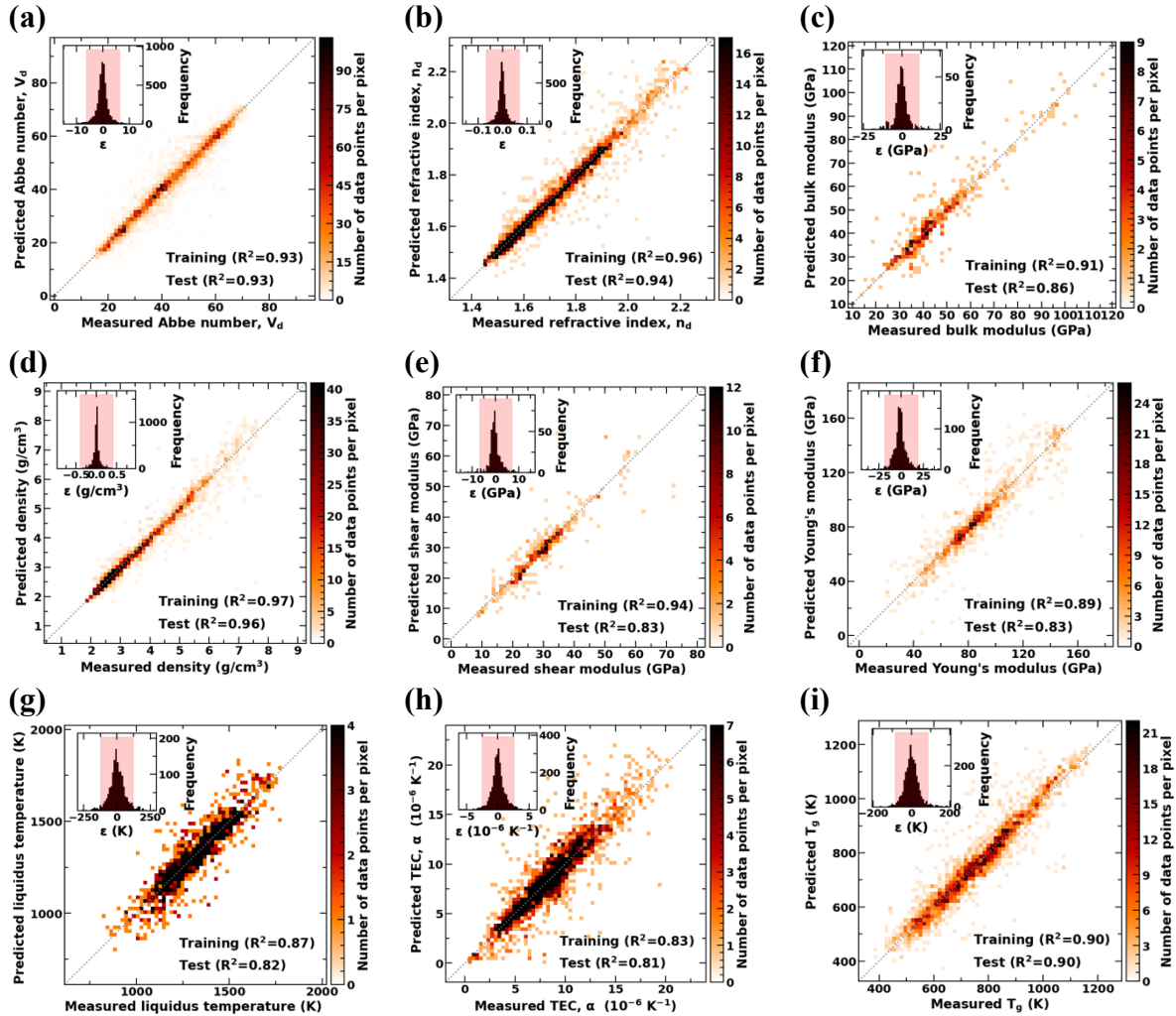

**Figure 1. ML model performance using LCNN.** Predicted values of (a) Abbe number, (b) refractive index, (c) bulk modulus, (d) density, (e) shear modulus, (f) Young's modulus (g) liquidus temperature, (h) thermal expansion coefficient, and (i) glass transition temperature with respect to measured values using LCNN and descriptors.

First, we focus on the performance of LCNN for predicting the properties of glasses using the descriptors as the input features. Figure 1 shows the predicted values based on the trained LCNN model on the unseen test dataset for the nine properties considered. We note that the $R^2$ values for all the models are greater than 0.8 suggesting good performance of the model on the test data. In addition, we observe that for properties such as Abbe number, density, refractive index and glass transition temperature, the $R^2$ values greater than 0.90, suggesting excellent generalizability of the model to unseen data. In order to evaluate the performance of LCNN in comparison to other ML models, we select XGBoost as recent studies have shown that tree-based approaches provide the best results for property prediction in small dataset and especially glasses. Table 2 shows the results for the obtained using the XGBoost models for the same test set after performing training using the state of the art methods and hyper-parametric optimization (see Methods). We observe that for most properties LCNN gives better performance that XGBoost. This suggests that the modification in the loss function indeed results in the improved performance of the model and better generalizability. The train, validation, and test set results of the LCNN models are provided in the Supplementary material.

Further, to evaluate the perform of LCNN on extreme regions with sparse data, we consider the highest and lowest 5% data for each property from the test data. As evident from Fig. 1 these regions have sparse data in comparison to the other regions. More importantly, for discovering new materials, researchers are mainly interested in this region. For instance, glasses with high hardness, or low Abbe number are common requirements. As such, it is important to have models with high-fidelity in these regions. Table 3 shows the $R^2$ values of the predictions on the highest and lowest 5% data for each property using LCNN and XGBoost. Interestingly, we also observe that predictions in the extreme regions by LCNN is significantly better than that by XGBoost for all properties consistently. This could be attributed to the weighted loss function, where the inverse of the number of datapoints is used to weight the localised losses after binning the data based on the y-values (see Methods for details). This approach enables models to be applied to discover new glass compositions with extreme property values with improved reliability, as evinced by improved predictions in extreme regions (low and high y-values).

Finally, we compare the performance of the models trained with composition as input. Interestingly, for five properties, namely, density, liquidus temperature, refractive index, thermal expansion coefficient, and glass transition temperature, we observe that the descriptor based LCNN model gives comparable or better performance than composition-based DNN models. This suggests that the descriptors chosen here can represent the features of the glasses that govern the composition–property relationships, while reducing the dimensionality of the problem significantly (34 to 12). For all the moduli, that is, shear, bulk, and Young's moduli, we observe that the descriptor-based models give slightly lower performance than the composition based models. This suggests that the inclusion of additional descriptors capturing the elastic deformation behavior can potentially improve the predictions for moduli.

| **Table 2.** Comparison of the $R^2$ values for the ML models with descriptor based features ($XGBoost_D$ and $LCNN_D$) and composition-based features on the test dataset. | | | |
|---|---|---|---|
| **Property** | **DNN (Composition as input)** | **Descriptor XGBoost** | **Descriptor LCNN** |
| Abbe number | - | 0.91 | **0.93** |
| Refractive index | 0.94 | 0.94 | **0.94** |
| Bulk modulus | 0.89 | 0.83 | **0.86** |
| Density | 0.95 | 0.96 | **0.96** |
| Shear modulus | 0.88 | **0.84** | 0.83 |
| Young's modulus | 0.86 | 0.83 | **0.83** |
| Liquidus temperature | 0.80 | **0.84** | 0.82 |
| Thermal expansion coefficient | 0.80 | 0.78 | **0.81** |
| Glass transition temperature | 0.90 | 0.88 | **0.90** |

**Table 3.** Comparison of the R² values for XGBoost and LCNN models on extreme region test data (highest 5% and lowest 5% data from the test).

| Property | XGBoost | LCNN |
|---|---|---|
| Abbe number | 0.955 | **0.964** |
| Refractive index | 0.930 | **0.941** |
| Bulk modulus | 0.911 | **0.953** |
| Density | 0.953 | **0.958** |
| Shear modulus | 0.845 | **0.865** |
| Young's modulus | 0.841 | **0.887** |
| Liquidus temperature | 0.901 | **0.904** |
| Thermal expansion coefficient | 0.763 | **0.848** |
| Glass transition temperature | 0.924 | **0.950** |

**Interpreting the ML models**

Traditional ML models such as NNs are black box in nature and hence do not provide insights into the role of each of the descriptors in governing the target properties. However, in understanding and rationalize the behavior of materials such insights are crucial. Thus, in order to gain insights into the descriptor–property relationships, we interpret the models using the Shapely additive explanations (SHAP) values. SHAP employs a game-theoretic model agnostic approach toward interpreting the input–output relationship in an ML model. Specifically, SHAP enables a quantitative analysis of whether each descriptor controls an output property positively or negatively, that is, it demonstrates whether increasing a particular descriptor value increases or decreases a given property and by how much.

For each property, the SHAP outcomes are displayed through two subplots, namely, river flow (top subplot, see Fig. 2(a)) and beeswarm (bottom subplot, see Fig. 2(a)). The river flow plot shows for a given composition what is the contribution of each descriptor towards the target output property. Thus, a line in the riverflow plot, from left to right, shows the evolution of the property value with each increase or decrease depicting the role of the specific descriptor. The final property value corresponding to each composition is shown in the right y-axis. The lines are thus colored according to the final property value with high value towards red and low values toward blue. In contrast, the beeswarm plot how each descriptor controls the output value. That is, whether increasing a descriptor value result in increase in the property or vice versa. In the beeswarm plot, the color represent the descriptor value while the y-axis represents the SHAP value, that is, the contribution towards the property value. Thus, while the riverflow plot describes the SHAP values with focus on a given composition, the beeswarm plot describes the SHAP values with focus on a given descriptor. It should be noted that the SHAP values are always with respect to the mean value of the property of the dataset. Thus, SHAP values need not be considered universal and quantitative. However, they are extremely useful in capturing the trend of each of the input feature towards the output value.

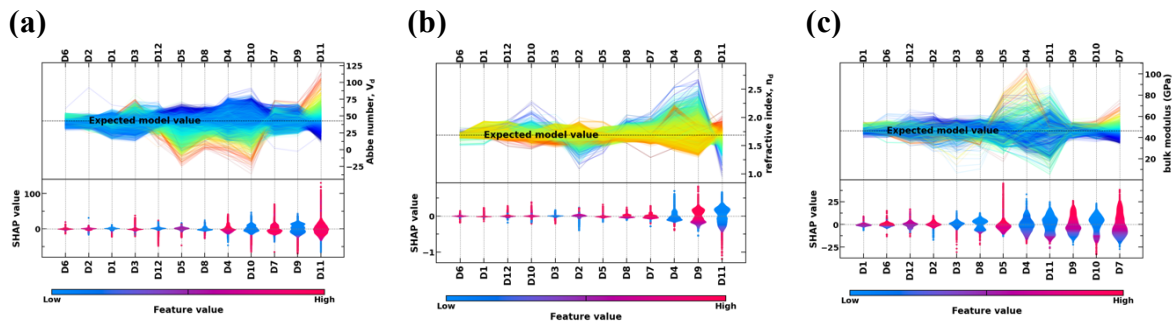

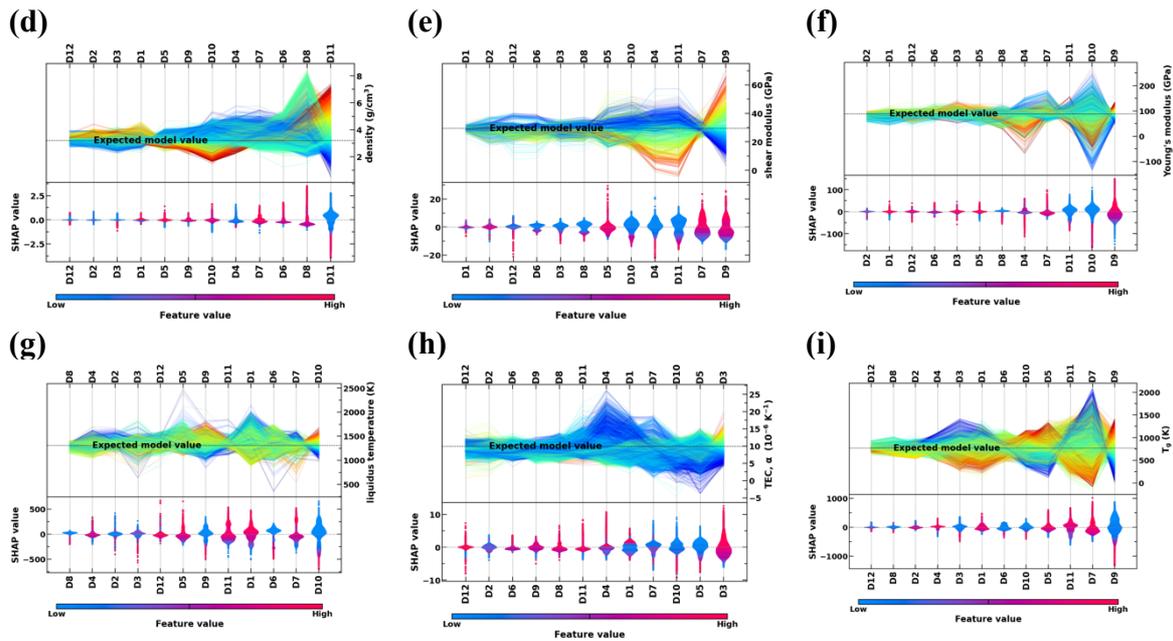

**Figure 2.** LCNN shap value plot for: (a) Abbe number, (b) refractive index, (c) bulk modulus, (d) density, (e) shear modulus, (f) Young's modulus (g) liquidus temperature, (h) thermal expansion coefficient, and (i) glass transition temperature.

Figure 2 shows the SHAP analysis of the final LCNN models. We first focus on the optical properties, namely, refractive index and Abbe number (see Fig. 2(a), (b)). Interestingly, all the features that enhance refractive index decreases the Abbe number and vice versa. This is also in agreement with the literature that any glass component that increases Abbe number decreases the refractive index (and vice versa) as revealed by the Abbe diagram. The top two features that govern optical properties are D11 and D9, that is, covalent and atomic radii, respectively. Interesting, we observe that while covalent radii enhance the Abbe number, atomic radii has a negative effect on it. Indeed an inverse trend is observed for refractive index. The valency of the network former (VNF) is the third important feature governing refractive index, although it is the fifth important feature for Abbe number. It is interesting to note that the percentage of oxygen, network former, and modifiers have little effect on the optical properties.

Now, we focus on the mechanical properties, namely, density, and Young's, shear and bulk moduli (see Fig. 2(c), (d), (e), (f)). Top four features governing density are covalent radii, atomic number, atomic mass, and atomic volume, respectively. Specifically, covalent radii has a negative effect while atomic number, atomic mass, and atomic volume has a positive effect. This is in agreement with the understanding that density of a glass is primarily governed by the mass of the species and their packing. The top five features governing the moduli are VNF, atomic volume, atomic, van der Waal and covalent radii, although in different orders for Young's, bulk, and shear moduli. While atomic volume has the maximum impact on bulk moduli, it is the second and fourth important features for shear and Young's moduli, respectively. Atomic radius has the maximum influence on shear and Young's moduli. Interestingly, features such as oxygen, network former, modifier percentage, and atomic mass has very little direct effect on the mechanical properties. Altogether, different radii of the species (atomic, covalent, van der Waal etc.) govern the mechanical properties of glasses to a large extent.

Now, we analyse physical properties, namely, liquidus temperature, glass transition temperature and the thermal expansion coefficient (TEC). Liquidus temperature is primarily governed by van der Waal's radius, atomic volume, atomic mass, and oxygen percentage, in the decreasing order of impact. While van der Waal's radius and atomic mass has a negative impact on the liquidus temperature, atomic volume and oxygen percentage has a positive impact on it. Top five features governing TEC are percentage of network modifiers, VNF, van der Waal radius, atomic volume, and oxygen percentage. Specifically, the network modifier increases the TEC, while the VNF, van der Waal radius, and atomic volume decrease the TEC. This observation is congruent with the fact that the network formers can decrease the TEC due to the increased polymerized structure, while network formers can enhance it. Finally, the top five descriptors governing $T_g$ are atomic radius, atomic volume, covalent radius, VNM, van der Waal radius. While atomic and van der Waal's radius has a negative effect on $T_g$, atomic volume, covalent radius, and VNM has a positive effect. Altogether, the SHAP analysis reveals the role of the descriptors in governing the glass properties. Many of these observations can be rationalized based on the existing knowledge. Further, some of these observations can also initiate additional research towards understanding the mechanism governing the descriptor–property relationships. The detailed SHAP plots and the interaction value plots of the XGBoost models are provided in the Supplementary Material.

**Universality of the ML models**
One of the major disadvantages of composition-based ML models is that they are restricted to the original set of components on which they are trained. Converting the compositions to the descriptor space can potentially address this problem, since the descriptors can be obtained for any chemical species or components. To test the universality of the ML models, we test the ability of these models to predict on glasses with unseen components. Note that this problem is harder than predicting the properties of unseen compositions but with the same components. Further, this test will also enable an evaluation on whether the models developed here are universal, that is, whether they can be used for the oxide of any element in the periodic table provided they are included in the glass.

To test the universality of the features, extracted new glass compositions from the literature for the properties trained here. Specifically, we focussed on binary, ternary, and quarternary glasses with one new component (that is, a component that was not present in the original training list) and the remaining being from the initial list of components. This was done to ensure that the new glass compositions have a similar trend as the ones in the training set, while having new components. Figure 3 shows the predictions from the LCNN models in comparison with the experimental values for density, refractive index, TEC, and $T_g$. For other properties, the data from literature for a new glass composition with one new component was not available. This is also because the original dataset considered for the study was quite exhaustive and comprised of all major oxide components. The legend for each subfigure shows the new component added in a square bracket and the other components that were already present in the training set. We observe that for most new glass compositions the predictions are fairly close to the experimental values. Indeed, we observe an increased spread in some family such as $In_2O_3$, however, for most cases, we observe that the models are able to extrapolate to new compositions that are beyond the training, suggesting the potential of these models to serve as universal property prediction models.

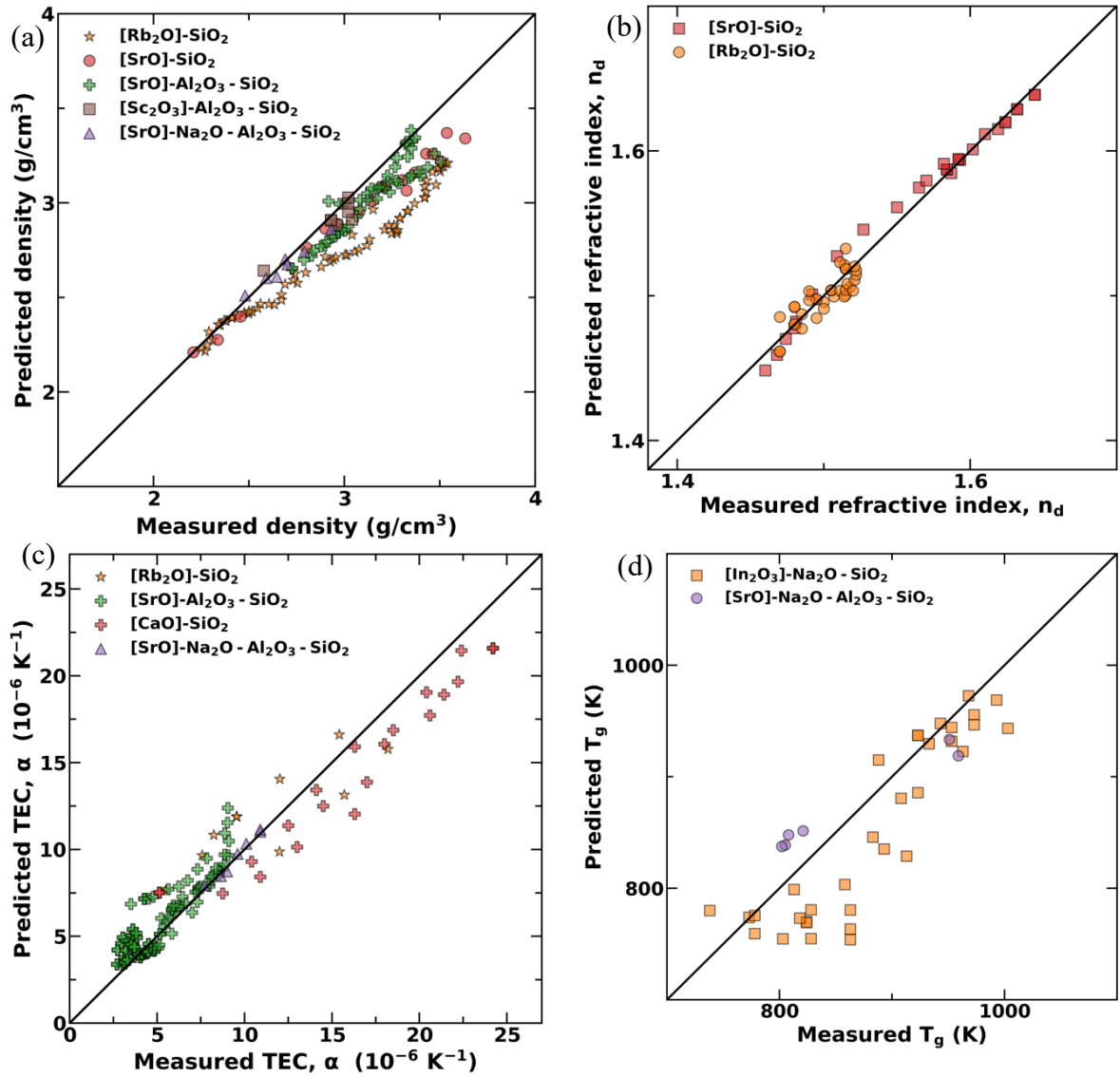

**Figure 3.** Predicted values of (a) density, (b) refractive index, (c) TEC, and (d) $T_g$ using the LCNN models with respect to the experimental values binary, ternary, and quarternary glass compositions with a new component that was not present in the original dataset. New component is shown by square brackets [] around the compound in the legend of the plot.

**Conclusions**

In conclusion, here, we presented a new LCNN framework incorporating physical and chemical descriptors for predicting and interpreting the composition–property relationship of oxide glasses. We showed that the LCNN framework provides better performance than the state-of-the-art methods such as XGBoost and also gives superior performance for extreme values, which are generally the regions of interest. In addition, we showed that the performance of LCNN is comparable to that obtained from the composition-based models establishing the ability of the descriptors to capture the compositional information. We interpreted the models employing SHAP analysis to decrypt the role of each of the descriptors. Finally, we demonstrated the universality of these models to make predictions for compositions with completely new components that were not present in the original dataset. Altogether, the present work paves the way towards developing universal models with improved accuracy that can enable accelerated materials design and discovery.

## Methodology
### Dataset preparation

Data used in this work collected from commercial glass database, namely, INTERGLAD© Ver.7. and is the same as previous works (Bishnoi et al., 2021; Ravinder et al., 2020). Data was extracted for nine properties, namely, Abbe number, bulk modulus, density, liquidus temperature, refractive index, shear modulus, thermal expansion coefficient, glass transition temperature and Young's modulus. Data cleaning for these properties was done by following standard approaches (Ravinder et al., 2020) which include (i) removing duplicates entries, that is, identifying duplicates for compositions and then taking mean of the property (within ±3σ range) for those duplicate entries, and (ii) checking sum of the compositions add up to 100% (with ±1% tolerance). The final glass compositions were made of 34 components, namely, $SiO_2$, $B_2O_3$, $Al_2O_3$, MgO, CaO, BaO, $Li_2O$, $Na_2O$, $K_2O$, $Ag_2O$, $Cs_2O$, $Tl_2O$, BeO, NiO, CuO, ZnO, CdO, PbO, $Ga_2O_3$, $Y_2O_3$, $La_2O_3$, $Gd_2O_3$, $Bi_2O_3$, $TiO_2$, $ZrO_2$, $TeO_2$, $P_2O_5$, $V_2O_5$, $Nb_2O_5$, $Ta_2O_5$, $MoO_3$, $WO_3$, $H_2O$, $Sm_2O_3$. These compositions were converted to the descriptors as discussed in the main manuscript and the updated dataset along with the properties were used for training the model.

### Extreme Gradient Boosted Decision Trees (XGBoost)

To obtain a baseline, we train the machine learning models using the XGBoost algorithm. XGBoost stands for extreme gradient boosting of tree-based machine learning models such as tree ensemble. It contains tool set for scalable end-to-end tree boosting system, sparsity-based algorithms, and justified weighted quantile sketch for efficient proposal calculation. XGBoost uses K additive function to predict the output as

$$\hat{y}_i = \sum_{k=1}^{K} f_k(x_i), f_k \in F$$

where K in number of regression tree (which uses classification and regression tree (CART) algorithm) in a tree ensemble, f is function in functional space F, F is set of all possible CARTs and $x_i$ is input feature vector for $i^{th}$ data point in the given dataset $D = \{x_i, y_i\}, i = \{1, 2, 3, \ldots, n\}$ where n is total number of data points. The tree ensemble is created by iteratively adding new regression trees (CARTs) to improve model accuracy. Finding all possible regression trees which improve model accuracy is impractical. Therefore, an optimal regression tree is created from a single node by iteratively adding branches. Adding branches stops if the allowable depth of a tree is reached or the number of samples at a splitting node is equal to one or equal to the minimum number of samples. Such, tree-based algorithms have been shown to give SOTA performance for prediction of glass properties in comparison to other algorithms(Cassar et al., 2021a; Ravinder et al., 2021).

### Low Complexity Neural Networks

The low complexity neural network (Jayadeva et al., 2021) (LCNN) is a feed-forward neural network with a modified loss function that consists of an upper bound on the model complexity as defined by the Vapnik-Chervonenkis (VC) dimension (Vapnik et al., 1994). Traditional NNs are typically trained by minimizing a loss function with respect to the network weights. This is done by back-propagating the gradient of the loss function, and moving weights in the direction of the negative gradient of the loss function. The loss function usually contains a regularization term, such as $|w|^2$, or the $L^2$ norm of the weights, as well as a term that penalizes the classification or the estimation error. In recent years, minimizing model complexity has gained vogue and newer approaches have been proposed (Han et al., 2015; Sharma et al., 2017). While the classification error is characterized by the model performance metrics such as F1 score or the confusion matrix, the model complexity can be expressed in terms of the VC dimension (Vapnik et al., 1994). The LCNN tries to learn a model with a small VC dimension, that also minimizes the classification error. Minimizing network complexity for classification problems was proposed in a previous work (Jayadeva et al., 2021) by using a loss function consisting of an upper bound on the VC dimension plus an error term. Here, we modify the LCNN for regression problems by using the complexity term along with a mean squared error term, as part of the loss function. Further, to account for the non-uniform distribution of the data training is performed with different number of bins (a hyperparameter) and the loss function is weighted with the inverse of the number of data points in the bin. The bins are formed by uniformly dividing the output range into $n$ equidistant regions. This way both regions with small and large number of data points gets trained equally well, which is demonstrated empirically in the results section. The LCNN loss function used for training the model is given by

$$Loss = \frac{c}{2} \sum_{i=1}^{n} \frac{\sum_{j=1}^{m_i} |\hat{y}_i^{(j)} - y_i^{(j)}|}{w_i} + \frac{1}{2} \sum_{k=1}^{N} (\hat{y}^{(k)})^2$$

where, $c$ is a hyperparameter, $y_i^{(j)}$ represents the output value of the $j^{th}$ datapoint in the $i^{th}$ bin, the (^) represents the predicted value, $n$ represents the number of bins in which the $y$ values are distributed, $N$ represents the total number of datapoints in the training set, $m_i$ represents the number of datapoints in the $i^{th}$ bin, and $w_i = \frac{m_i}{N}$. The $w_i$ in the first term gives increased weightage to regions with lower datapoints so that the model learns function even in regions where the dataset is sparse.

**Model development and hyperparameter optimization**
Different ML models were developed for each of the Nine properties. We use squared error as our loss function for optimizer. We use two types of boosters namely GBtree and DART. Further, we use Optuna to do hyperparameter optimization for XGBoost tree ensemble models. The range for the hyperparameters is given in the Tables 4, 5, and 6. Finally, the model with the best validation score is selected.

| Table 4. XGBoost tree ensemble specific hyperparameters | |
|---|---|
| **Name of hyperparameter** | **Range/List/Value** |
| booster | dart, gbtree |
| lambda | 1e-8 to 1.0 in log scale |
| alpha | 1e-8 to 1.0 in log scale |
| Number of estimators | 300 |
| subsample | 0.7 to 1 |
| colsample_bytree | 0.7 to 1 |
| reg_alpha | 1e-4 to 1 in log scale |
| reg_lambda | 1e-4 to 1 in log scale |

| Table 5. Booster (DART and GBtree) specific hyperparameters | |
|---|---|
| **Name of hyperparameter** | **Range/List/Value** |
| max_depth | 1 to 9 |
| Eta | 1e-8 to 1.0 in log scale |

| Gamma | 1e-8 to 1.0 in log scale |
| grow_policy | depthwise, lossguide |

| Table 6. DART specific hyperparameters | |
|---|---|
| **Name of hyperparameter** | **Range/List/Value** |
| sample_type | uniform, weighted |
| normalized_type | tree, forest |
| rate_drop | 1e-8 to 1.0 in log scale |
| skip_drop | 1e-8 to 1.0 in log scale |

**Hyperparametric optimization**

Hyperparametric optimization was performed using optuna, an open-source software package. It provides a define-by-run programming environment, efficient sampling and pruning algorithms, modular and is easy to scale. Optuna optimizes (minimize or maximize) an objective function which takes a set of hyperparameters (called trial) as an input and returns the validation score as output. It uses both relational (namely CMA-ES and GP-BO) and independent (TPE) sampling algorithms to investigate new trials. It also uses pruning techniques like Asynchronous Successive Halving (ASHA) for termination of inefficient trials. It also provides an interface using which we can provide custom sampling and pruning methods. Finally, we get an optimized set of hyperparameters and trained ML model.

**Shapley additive explanations for tree ensemble**

Shapley additive explanation (SHAP) values is a unified game theoretic approach to calculate the feature importance of an ML model. SHAP measures a feature's importance by quantifying the prediction error while perturbing a given feature value. If the prediction error is large, the feature is important, otherwise the feature is less important. It is an additive feature importance method which produces unique solution while adhering to desirable properties namely local accuracy, missingness, and consistency. SHAP introduces model agnostic approximation methods such as Kernel SHAP as well as model-specific approximation methods such as Deep SHAP, Max SHAP, and Linear SHAP, etc. Here, we use a tree-specific SHAP approximation method namely TreeExplainer. Shapley value calculation requires a summation over all possible feature subsets, which leads to an exponential time complexity. TreeExplainer exploits the internal structure of tree-based models and collapse summation to a set of calculation specific to the leaf node of a tree model leading to low order polynomial time complexity. TreeExplainer also calculates interaction values to capture the local interaction effects. These local interaction values can be used to explain the effects of feature interaction.